\documentclass[useAMS,usenatbib]{mn2e}
\usepackage{graphicx}
\usepackage{epsfig}
\usepackage{color}

\def\la{\mathrel{\hbox{\rlap{\hbox{\lower4pt\hbox{$\sim$}}}{\raise2pt\hbox{$<$}}}}}
\def\ga{\mathrel{\hbox{\rlap{\hbox{\lower4pt\hbox{$\sim$}}}{\raise2pt\hbox{$>$}}}}}

\title[Spectral variability in the ULX Holmberg IX X-1]{X-ray spectral variability in the ultraluminous X-ray source Holmberg IX X-1}
\author[K. Vierdayanti, C. Done, T.\,P. Roberts, and S. Mineshige]{K. Vierdayanti$^{1}$\thanks{E-mail:kiki@kusastro.kyoto-u.ac.jp}, C. Done$^{2}$, T. P. Roberts$^{2}$, and S. Mineshige$^{1}$\\
$^{1}$Department of Astronomy, Kyoto University, Kitashirakawa Oiwake-cho, Kyoto  6068502, Japan\\
$^{2}$Department of Physics, University of Durham, South Road, Durham DH1 3LE, United Kingdom}
\begin{document}

\date{Accepted 2009 December 00. Received 2009 December 00; in original form 2009 October 00}

\pagerange{\pageref{firstpage}--\pageref{lastpage}} \pubyear{2009}

\maketitle

\label{firstpage}

\begin{abstract}
We use {\it XMM-Newton\/} and {\it Swift\/} data to study spectral
variability in the ultraluminous X-ray source (ULX), Holmberg IX
X-1. The source luminosity varies by a factor 3 -- 4, giving rise to
corresponding spectral changes which are significant, but subtle, and
not well tracked by a simple hardness ratio. Instead, we co-add the
{\it Swift\/} data in intensity bins and do full spectral fitting with
disc plus thermal Comptonisation models. All the data are well-fitted
by a low temperature, optically thick Comptonising corona, and the
variability can be roughly characterised by decreasing temperature and
increasing optical depth as the source becomes brighter, as expected
if the corona is becoming progressively mass loaded by material blown
off the super-Eddington inner disc.  
This variability behaviour is seen in other ULX which have similar spectra,
but is opposite to the trend seen in ULX with much softer spectra.
This supports the idea that there are two distinct physical regimes in ULXs,
where the spectra go from being dominated by a disc-corona to being dominated
by a wind.

\end{abstract}


\section{Introduction}

Ultraluminous X-ray sources (ULXs) are X-ray sources that are too
bright ($L_{\rm X}>10^{39}$~ergs~s$^{-1}$) to be powered by
sub-Eddington flows onto a stellar remnant black hole, but are not
associated with galactic nuclei so are not powered by accretion onto a
central supermassive black hole.  If the Eddington limit holds, then
the central object must be intermediate in mass between the population
of $\la 20 M_\odot$ stellar remnant black holes known in our Galaxy
and the supermassive black holes which form in the centres of galaxy
bulges, and are therefore termed intermediate-mass black holes
(IMBHs). However, there are many serious problems with such an
interpretation, not least the {\em in situ} formation of the large
numbers of such objects required to explain the populations of ULXs in
star forming regions, that would result in an unfeasibly high fraction
of mass being held in the IMBHs \citep{b37}. Thus the majority of ULXs
are most probably stellar remnant black holes\footnote{It is possible
that stellar remnant black holes may range up to $\sim 90 M_{\odot}$
in mass if formed in low-metallicity environments, see \cite{b105}.},
accreting at super-Eddington rates (e.g. King et al. 2001; Watarai,
Mizuno \& Mineshige 2001; Begelman 2002; see also Roberts 2007 and
references therein), though in the absence of a direct dynamical mass
measurement we cannot yet confirm this for any individual ULX.
Possibly the best current IMBH candidates amongst ULXs are those with
the highest observed luminosities, and those with mHz QPO detections
(although see below on the latter point).  However, even amongst the
brightest ULXs (the hyperluminous X-ray sources, or HLXs, e.g. Gao et
al. 2003) there may be explanations other than an IMBH - for example,
the HLX reported by Miniutti et al. (2006) has subsequently been
revealed to be a background QSO (G. Miniutti, private communication),
and the record-breaking HLX reported by Farrell et al. (2009) may turn
out to be the stripped nucleus of a dwarf galaxy (Soria et al. 2009).

The bulk of the ULX population therefore probably probes a new mode of
accretion onto stellar-remnant black holes, different to the
sub-Eddington states that are well studied from observations of black
hole binaries (BHBs) in our own Galaxy (see e.g. the reviews by
Remillard \& McClintock 2006 hereafter RM06; Done, Gierlinski \&
Kubota 2007 hereafter DGK07). Thus they can constrain models of
super-Eddington accretion flows, which have wider impact than just
ULXs. In the local Universe, these flows power objects such as the
unique BHB GRS 1915+105, and the puzzling narrow line Seyfert 1s. They
also have cosmological significance as super-Eddington flows are
required in the early Universe in order to quickly build up black hole
mass to form the first quasars. ULXs can then give a template for
understanding super-Eddington accretion flows, in a similar way to
that of the BHBs for sub-Eddington flows.

\begin{figure*}
\includegraphics[angle=270,scale=0.65]{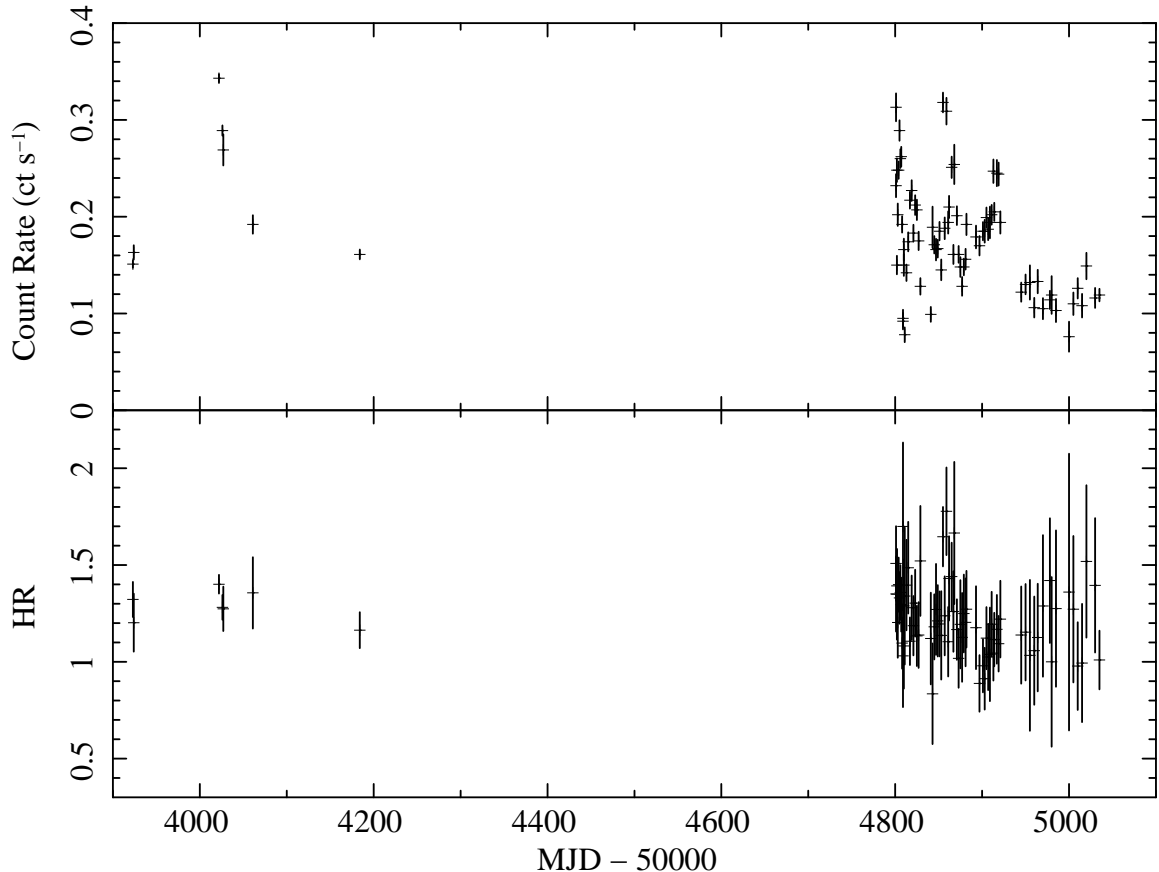}
 \caption{The {\it Swift\/} light curve and hardness ratio for
 Holmberg IX X-1.  We show the 0.3 -- 10 keV count rate as a function
 of observation time (top panel), and the corresponding hardness ratio
 (HR, as defined in the text; bottom panel).}
\label{figure1}
\end{figure*}

For BHBs the extensive, high quality monitoring data available means
that their spectral variability as a function of luminosity can be
studied for individual objects as well as for the class as a
whole. This has given rise to a well defined picture, where at low
luminosities the spectra are dominated by a hard $\Gamma \la 1.8$
power-law tail extending out to $\sim 100$~keV. The thermal emission
from an optically thick disc starts to become more important,
increasing in both temperature and luminosity as the source
brightens. This is correlated with a softening of the power-law tail,
which pivots towards $\Gamma\sim 2.1$ (low/hard state: LHS). The
spectrum then makes an abrupt transition to the high/soft state (HSS)
where the disc dominates the spectrum, though this is generally
accompanied by a weak non-thermal tail to high energies with $\Gamma
\ga 2.2$. In this state the disc varies with $L\propto T^4$ as
expected for a constant inner disc radius.  However, at these high
luminosities the non-thermal tail can also become strong, carrying a
large fraction of the power, at which point the disc spectrum is
strongly distorted by this Comptonisation (very high or steep
power-law state: VHS) see e.g. the reviews by RM06; DGK07.

By contrast, the much lower flux from ULXs means instead that there
are typically only snapshot observations of individual objects, so
putting together a picture of the spectral variability is much more
difficult. Nonetheless, this has been attempted, though the limited
quality of the data mean that often the spectra can be equally well
described by power-law or disc models, which obviously confuses the
spectral state identification (e.g. Gladstone \& Roberts
2009). \citet{b40} used two ASCA observations of IC 342 to track the
spectral change of the two ULXs in this field.  Both ULXs show a
transition between a clearly curved spectrum which can be fit with a
disc model, to one which can be fit by a power-law. Previous work has
interpreted such changes as being from the HSS to LHS, and as this
transition generally occurs at sub-Eddington luminosities then this
would favour an IMBH. Kubota et al. (2002) instead showed that this
could be interpreted as the much higher luminosity HSS to VHS
transition, consistent with a stellar remnant black hole. `Spectral
transitions' from sparse observations of other ULXs have also been
discovered \citep{b11,b12,b13,b14,b15}, but the results are confusing,
for example some ULXs have spectra which become harder as they become
brighter, while others show the opposite trend.  Recent work by
\cite{b101} and \cite{b102} has begun to systematically analyse {\it
XMM-Newton\/} data for individual ULX variability patterns, and
interpret it in a physical framework.

Another approach is to use higher signal-to-noise data from many {\em
different} ULX to try to delineate the spectral changes as a function
of luminosity, though this is complicated by the fact that ULXs are
likely to be a heterogeneous population of X-ray sources, in terms of
black hole mass and viewing angle. Generally these spectra are fit
with disc plus power-law models, and the derived disc temperatures are
sometimes low, $\sim$ 0.2~keV, which (if taken at face value) implies
a high black hole mass of $\sim$ 10$^3$ $M_\odot$ (e.g. Miller et
al. 2003; Kaaret et al. 2003; Miller, Fabian \& Miller 2004). While
this initially supports the IMBH LHS interpretation, these higher
quality data also clearly show that the accompanying power-law tail
curves at the highest energies, with a deficit of photons above 5~keV
(Stobbart, Roberts \& Wilms 2006; Miyawaki et al. 2009).  Such
curvature is never seen in the BHB LHS at these low energies, arguing
against a simple IMBH interpretation for these objects on the basis of
their X-ray spectra.  On a similar theme, the few tens of mHz QPOs
seen in some ULXs imply IMBHs if the observed QPOs are the direct
analogues of the strong low frequency type-C QPOs seen in BHBs at
1-10~Hz (e.g. Strohmayer \& Mushotzky 2009).  However, we note that
the likely super-Eddington, $\sim 14 M_\odot$ BHB GRS 1915+105 can
show QPOs at similarly very low frequencies (Morgan, Remillard \&
Greiner 1997, Hoel, Vaughan \& Roberts 2009).  Such QPOs are then
clearly not an unambiguous signature of IMBHs.

The highest quality X-ray spectra are actually much better fitted by
disc plus low temperature, optically thick Comptonisation models
(Stobbart et al. 2006; Roberts 2007; Gladstone, Roberts \& Done 2009).
Using this description, \citet{b7} sorted the 12 best available {\it
XMM-Newton\/} ULX spectra into a possible luminosity sequence,
starting from a disc dominated, bright high/soft state at around the
Eddington limit. For higher mass accretion rates they suggested an
increasing fraction of the power is dissipated in a corona which
covers the inner disc. This corona increases in optical depth and
decreases in temperature as the source brightens, perhaps due to the
increase in mass loss from the inner disc for these super-Eddington
flows. At the highest luminosities the corona is strongly outflowing,
giving rise to an increasingly large photosphere around the source
which thermalises the energy to progressively lower temperatures
(e.g. Kawashima et al. 2009; Begelman, King \& Pringle 2006; Poutanen
et al. 2007).

While plausible, such models must be tested on individual objects
where the mass and inclination of the black hole are fixed since ULXs
seem likely to be a heterogeneous population, as stressed above.  {\it
SWIFT\/} has been conducting a monitoring program for some ULXs in
recent years \citep{b8,b9}. In our present study, we use the data
obtained from the most intensive of these, that of Holmberg IX X-1
(also historically known as M81 X-9).  This source is located close to
the dwarf galaxy Holmberg IX ($d =3.4$ Mpc) and is the brightest X-ray
source within 20 arcminutes of the nucleus of M81 \citep{b17}.  This
is also one of the ULXs that has been most widely studied in the past
ten years \citep{b17,b18,b31,b32,b33,b9}. However, all these either
used hardness ratios to follow the spectral variability, or at best
fit disc plus power-law models. Here we choose to follow the spectral
variability of Holmberg IX X-1 from the {\it Swift\/} monitoring data,
but model this in detail using the disc and low temperature, optically
thick Comptonisation which is required to fit the highest quality {\it
XMM-Newton\/} data from this source \citep{b25,b7}. Section 2
describes how we extract and bin the intensity sorted spectra, while
Section 3 gives details of the spectral fitting results. Section 4
describes how these broadly fit into the \citet{b7} picture of
spectral variability described above, but also shows the need for
better data to understand these trends.

\section[]{Data}

We use only those {\it Swift\/} XRT observations that are pointed at
Holmberg IX X-1, so we can directly compare instrument count rates
rather than having to correct for the different instrument response of
off-axis pointings. This gives a total of 75 pointings, each typically
with $\sim 1700$~s exposure, from {\it Swift\/} program 00090008,
covering the whole monitoring program of \citet{b9} up to July
2009. In addition, we also use 7 observations from {\it Swift\/}
program 00035335 during 2006 -- 2007. These are longer observations
(up to 20ks) but have a much sparser coverage of the light curve.

We extract the photon counting mode of the {\it Swift\/} pointed
observation data using the {\it Swift\/}-specific \textsc{ftool} {\tt
xrtpipeline} to obtain clean event files. We use standard data
selection criteria, with minimum elevation angle (ELV) set to be
larger than 45$^{\circ}$, the bright Earth (BR$\_$EARTH) angle set to
be above 120$^{\circ}$, minimum Sun and Moon angle (SUN$\_$ANGLE and
MOON$\_$ANGLE) above 45$^{\circ}$ and 30$^{\circ}$, respectively, and
pointing direction within 0.08$^{\circ}$ of the source. We then
extract spectra for both source and background from grades 0 -- 12
within a 47 arcsec (20 pixels) radius region centred on the ULX.  A
background spectrum was extracted from a same-sized aperture on the
edge of the field-of-view.

Fig. 1 (upper panel) shows the resulting light curve for these data
over the 0.3 -- 10~keV band. Plainly the source is persistent, but
variable by a factor of 3 -- 4 in intensity. We also show the hardness
ratio (HR) for these data, defined using the standard {\it Swift\/}
XRT hard and soft energy bands as the rate from 1.5 -- 10~keV divided
by the rate from 0.3 -- 1.5 keV. The lower panel on Fig. 1 shows that
this HR does vary, but with large uncertainties associated with the
short observations.  We show the variation of HR with intensity in
Fig.  2, where the 7 data points from the 2006 -- 2007 observations
are shown in (magenta) filled stars while those from the monitoring
program are shown as (cyan) crosses.  This shows that the spectrum
hardens slightly as the luminosity increases, as seen in the analysis
of these data by \citet{b9} (although their hardness ratio is defined
slightly differently, as a ratio of 1 -- 8~keV/0.3 -- 8~keV).

\begin{figure}
 \includegraphics[angle=0,scale=0.45]{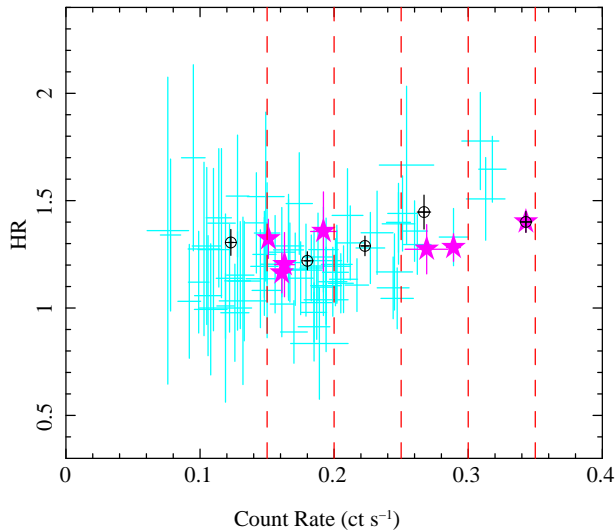} 
\caption{The hardness ratio-intensity diagram for Holmberg IX X-1.
We show each individual observation data point from the monitoring
programme as a cyan cross, while the data points from the earlier,
longer observations are indicated by magenta filled stars.  The count
rate ranges used for co-adding the data are delineated by the red
dashed vertical lines, and the values from these co-added data are
shown by black circles.}
\label{figure 2}
\end{figure}

We co-add observation data in order to increase signal-to-noise.  We
define 5 count rate ranges as detailed in Table 1 to have roughly
(within a factor 3) equal numbers of total counts as shown by the
(red) dotted lines on Fig. 2. Data within these ranges are consistent
with the same HR except for SWIFT 5, the highest intensity bin, where
the long observation in 2006  has a slightly but significantly
softer spectrum than the ones from the monitoring campaign. These 
data cannot be co-added as their spectra are not consistent. Hence we
include the 2006 data only for this intensity bin as this has higher
statistics.

\begin{table*}
 \centering
\begin{minipage}{160mm}
 \caption{The table details the datasets we use from \emph{Swift} to
 make the co-added spectra with higher signal-to-noise and
 \emph{XMM-Newton}. The prefix M and L on \emph{Swift} denotes
000900080 and 000353350, respectively.}
\label{bigtable}
\begin{tabular}{@{}lcrlllllll@{}}
\hline

Name & count rate & total counts & $\ \ \ \ \ \ \ \ \ \ \ \ \ \ \ \ \
\ \ \ \ \ \ \ \ \ \ \  $  \emph{Swift} and \emph{XMM-Newton} Obsid \\
\hline
SWIFT 1 & 0.00-0.15  & 4882 & M04, M12, M13, M14, M17, M18, M26, M28, M34, M45, M46, M47, M65, M66, \\
& & & M67, M68, M69, M70, M71, M72, M73, M75, M76, M77, M78, M79, M81, M82.\\
SWIFT 2 & 0.15-0.20 & 11733 & M11, M15, M19, M22, M25, M29, M30, M31, M32, M33, M36, M38, M41, M44, \\
& & & M48, M49, M50, M52, M54, M55, M56, M57, M58, M64, L01, L02, L07, L09. \\
SWIFT 3 & 0.20-0.25 & 7143 & M02, M03, M05, M07, M20, M21, M23, M24, M39, M43, M59, M60, M61, M62, \\
& & & M63.\\
SWIFT 4 & 0.25-0.30 & 6551 & M08, M09, M10, M40, M42, L05, L06.\\
SWIFT 5 & 0.30-0.35 & 6773 & L04\\
XMM1   &$1.53 \pm 0.005$ & 114000 & 0200980101 \\
XMM2   &$1.80 \pm 0.02$ & 12800 & 0112521001 \\
XMM3   &$2.07 \pm 0.02$ & 16000 & 0112521101 \\
\hline\\
\end{tabular}
\end{minipage}
\end{table*}

We add the spectra of each count-rate group using the {\tt addspec}
\textsc{ftool}. As recommended, the response files are weighted
with their counts. The differences between the auxiliary response
files ({\tt *.arf}) over all data sets in each count-rate range is
less than 10\%.
We include the hardness ratios from these co-added spectra on the
hardness ratio-intensity plot (Fig. 2) as
the large open circle (black) points.  For the purposes of spectral
fitting, the data were binned with at least 20 counts in each bin.

We also include the long exposure (114ks) {\it XMM-Newton\/} data of
Holmberg IX X-1 from \citet{b7}, hereafter termed XMM1 (observation ID
0200980101, taken on 2004-09-26). The spectrum of this {\it XMM-Newton\/}
data is reduced by using the same method as in \citet{b7}. These data have
very similar 0.3 -- 10~keV flux to that of the lowest luminosity {\it
Swift\/} spectrum.  However, its hardness ratio (when calculated
independently from the instrument response using model fluxes in the
bands) is significantly softer, closest to that from SWIFT 4. This
clearly shows that the variability is not simply a function of
luminosity, and that there is extra complexity present.

There are two other {\it XMM-Newton\/} data sets available in the
archive (hereafter XMM2 and XMM3).  These are all between XMM1 and
SWIFT 2 in intensity, so do not span much of a range in flux.  They
also span the range in shape between XMM1 to SWIFT 2, so do not add to
the spectral variability but we include them for
completeness.

\section{Model fitting and results}
 
\begin{figure*}
 \includegraphics[angle=0,scale=0.47]{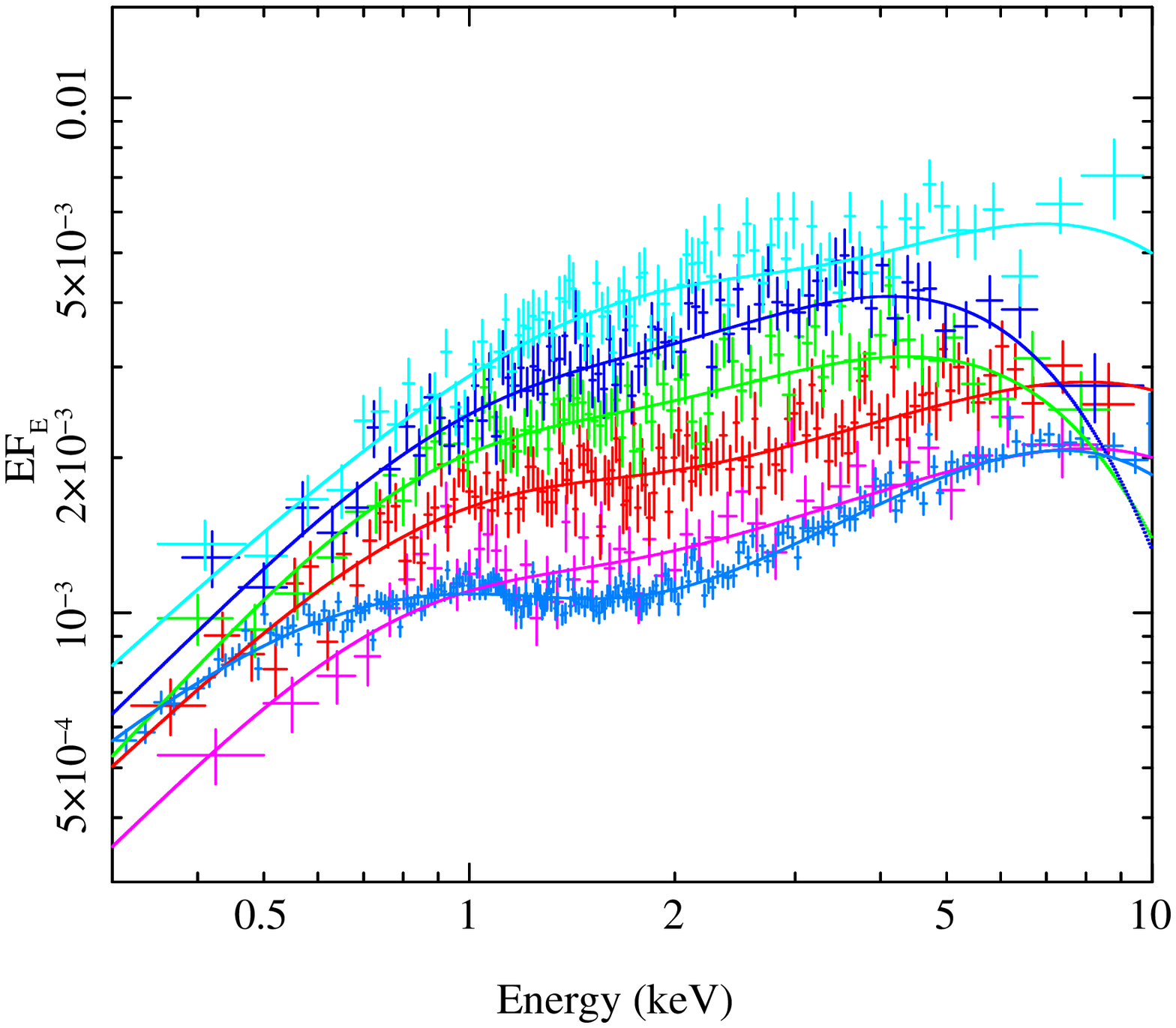}
 \includegraphics[angle=0,scale=0.47]{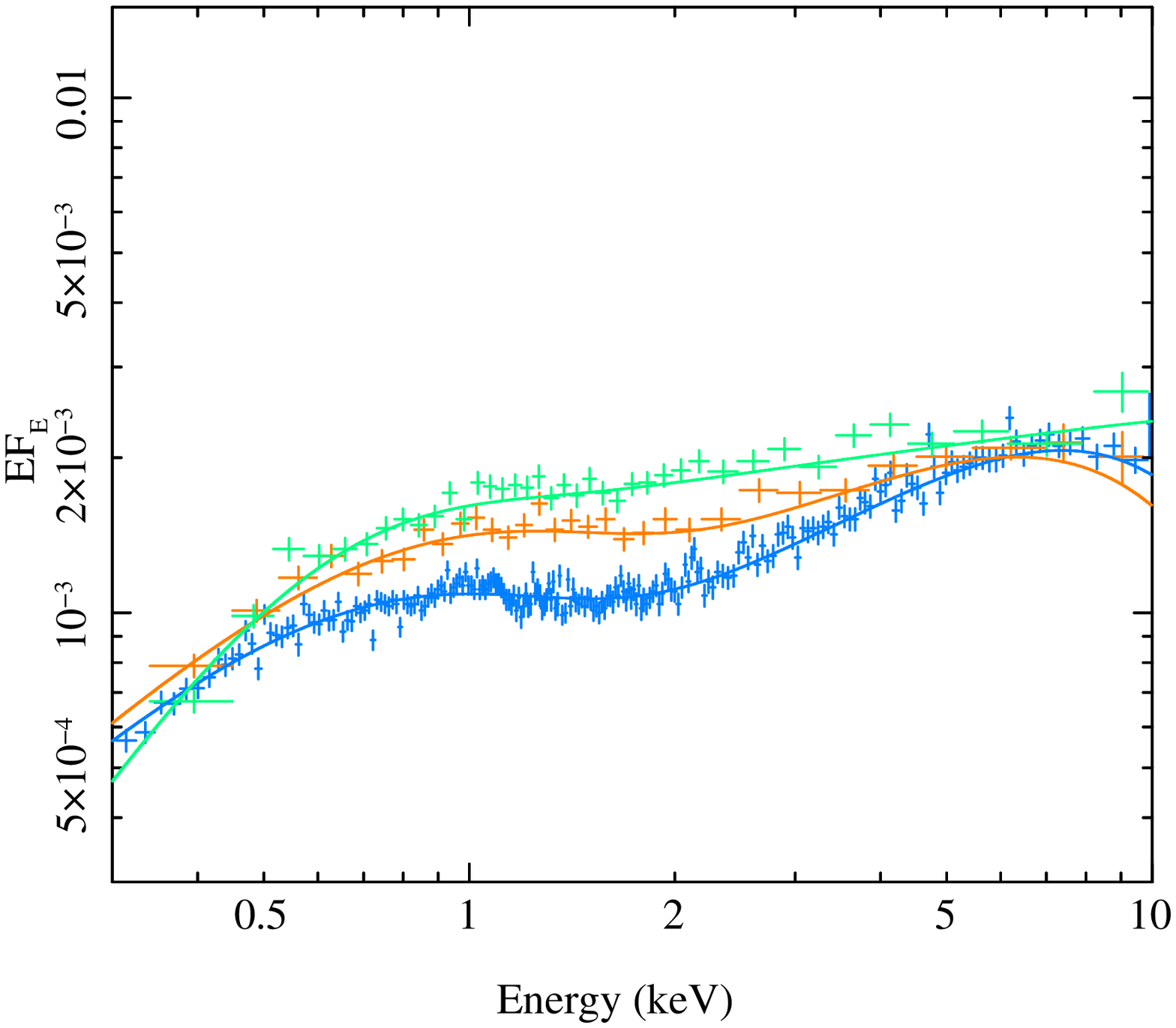}
\caption{Spectral data and corresponding best fitting model.
The data are de-absorbed and unfolded from the detector response.
Left panel from bottom to top: XMM1 (light blue), SWIFT 1 (magenta),
SWIFT 2 (red), SWIFT 3 (green), SWIFT 4 (blue), and SWIFT 5 (cyan),
respectively. Right panel from bottom to top: XMM1 (light blue), XMM2 (orange),
XMM3 (light green).
The SWIFT data groups are plotted to have at least
8$\sigma$ significance
or are grouped in sets of 10 bins, while the {\it XMM-Newton\/} data
is plotted to a minimum 20$\sigma$ significance or 20 bins for
clarity.}
\label{figure 3}
\end{figure*}

In some earlier studies, XMM1 data was conventionally analysed
using the 'standard' disc plus power-law model (e.g. Dewangan et al. 2006;
Winter, Mushotzky \& Reynolds 2007).
While this conventional model provides a reasonably acceptable fit,
the spectrum curves at high energy such that a power-law tail overpredicts
the observations above 5~keV \citep{b7}. Instead, a
disc plus low temperature (and hence optically thick) Comptonisation
spectrum fits this spectral shape, as the low electron temperature
gives the low energy rollover in the Comptonised flux. Hence we use
this model to fit all our spectra. \citet{b7} used some very
sophisticated disc and Comptonisation modelling to fit their sample containing
the longest exposure {\it XMM-Newton\/} ULX data available in the archive.
Here we have more limited
signal to noise in the {\it Swift\/} spectra so we choose the simplest
possible model for disc plus low temperature Comptonisation, namely a
{\tt diskbb} plus {\tt comptt} model, and set the seed photon
temperature equal to the disc temperature.

We assume a HI column density of 4.06$\times 10^{20}$ cm$^{-2}$ in the
direction of Holmberg IX X-1 \citep{b35}, and we allow intrinsic
absorption in the host galaxy to be a free parameter. We fit our 8
spectra (3 {\it XMM-Newton\/} plus 5 {\it Swift\/} spectra)
simultaneously in \textsc{xspec}, allowing all parameters to be free
except for the extra-galactic absorption column which is tied across
all 8 datasets.  This gives a reasonable fit, with
{\bf $\chi^2_\nu=3512/3461$} for a best fit value of
$N_H=(1.04 \pm 0.02)\times 10^{21}$ cm$^{-2}$. The derived disc and
Comptonisation parameters are given in Table 2, where the fluxes 
are the unabsorbed (intrinsic) flux from the 
{\tt diskbb} plus {\tt comptt} model from 0.3 -- 10 keV.
The error of the flux is calculated by using {\tt flux err} command in
\textsc{xspec}. The bolometric luminosity is calculated from the
unabsorbed flux in 0.01 -- 100 keV range, assuming disc geometry.

All the spectra are well fitted by a low temperature Comptonising
corona, that is optically thick.  However, the electron temperature
only has a firm upper limit in XMM1, XMM2, SWIFT 3 and SWIFT 4, so in the
other datasets the rollover is not significantly detected.
Nonetheless, the fact that some of the data (especially XMM1, see
Fig. 5 of Gladstone et al. 2009) do require this rollover shows that
we should use this model for all the data for consistency. Similarly,
the moderate signal-to-noise of the SWIFT data means that the disc
component is not strongly required in any of them ($\Delta \chi^2 \la
7$ for one additional degree of freedom) but we again include it for
consistency.

\begin{table*}
 \centering
 \begin{minipage}{133mm}
  \caption{Spectral fitting results. The flux is in units of $10^{-12}$ erg cm$^{-2}$ s$^{-1}$ while the bolometric luminosity, $L_{\rm bol}$, is in 10$^{39}$ erg s$^{-1}$. See text for the range of count rate for each SWIFT data group.  The errors quoted here and elsewhere in the paper are the 90\% confidence interval for one interesting parameter.}
  \begin{tabular}{@{}lccccccc@{}}
  \hline
   Group & $kT_{\rm in}$ & Norm.$^{1/2}$ & $kT_{\rm e}$ & $\tau$ & Flux\footnote{Unabsorbed flux in the fitting energy range (0.3 - 10 keV).} & $L_{\rm bol}$\footnote{The bolometric luminosity is calculated from unabsorbed flux in 0.01 - 100 keV range.} & $\chi^{2}_\nu$ \\
 \hline
SWIFT 1 & 0.203$^{+0.071}_{-0.069}$ & 5.04$^{+1.77}_{-2.52}$ & 3.01$^{+72.7}_{-1.11}$ & 6.86$^{+3.75}_{-4.42}$ & 6.22$^{+0.480}_{-4.47}$ & 6.63 & 189.6/171 \\ [3ex]
SWIFT 2 & 0.239$^{+0.175}_{-0.050}$ & 4.94$^{+1.13}_{-1.15}$ & 3.11$^{+58.7}_{-0.828}$ & 6.60$^{+4.31}_{-5.04}$ & 8.70$^{+0.640}_{-6.55}$ & 9.27 & 337.8/327 \\ [3ex]
SWIFT 3 & 0.215$^{+0.105}_{-0.059}$ & 5.47$^{+1.24}_{-2.73}$ & 1.68$^{+0.340}_{-0.223}$ & 9.16$^{+1.16}_{-1.17}$ & 9.36$^{+0.940}_{-2.73}$ & 9.41 & 226.1/245 \\ [3ex]
SWIFT 4 & 0.252$^{+0.359}_{-0.085}$ & 5.06$^{+1.79}_{-2.36}$ & 1.47$^{+0.221}_{-0.161}$ & 10.5$^{+2.67}_{-1.25}$ & 12.8$^{+0.600}_{-1.80}$ & 11.1 & 232.1/236 \\ [3ex]
SWIFT 5 & 0.488$^{+0.392}_{-0.241}$ & 2.35$^{+3.85}_{-0.964}$ & 2.47$^{+38.4}_{-0.813}$ & 8.15$^{+42.2}_{-6.32}$ & 18.0$^{+1.30}_{-14.7}$ & 17.3 & 227.1/244 \\ [3ex]
XMM1 & 0.237$^{+0.011}_{-0.05}$ & 5.67$^{+0.434}_{-0.395}$ & 2.44$^{+0.158}_{-0.140}$ & 8.76$^{+0.489}_{-0.426}$ & 5.95$^{+0.270}_{-0.570}$ & 6.45 & 1349.3/1286 \\ [3ex]
XMM2 & 0.263$^{+0.077}_{-0.053}$ & 4.98$^{+2.14}_{-1.13}$ & 2.31$^{+0.901}_{-0.309}$ & 8.09$^{+1.66}_{-1.82}$ & 6.65$^{+0.460}_{-2.65}$ &  6.92 & 435.7/447 \\ [3ex]
XMM3 & 0.159$^{+0.056}_{-0.027}$ & 6.43$^{+1.30}_{-3.21}$ & 9.67$^{+490}_{-6.61}$ & 3.07$^{+3.91}_{-2.95}$ & 7.91$^{+0.890}_{-6.23}$ & 11.1 & 514.2/506 \\ [1ex]
\hline
\end{tabular}
\end{minipage}
\end{table*}

Fig. 3, left panel, shows the derived spectra for XMM1 and the SWIFT spectra,
after correcting for absorption in both our Galaxy and the host galaxy, and
deconvolving from the instrument response. XMM2 and XMM3 are not
included on this plot as they overlap with XMM1, SWIFT1 and
SWIFT2. Instead, the three XMM spectra are shown separately in the right panel.

These plots clearly show the complex nature of the spectral variability. XMM1 
has a marked inflection at 1 -- 2~keV, separating the emission into two peaks 
which are fit by the cool disc and low temperature Comptonisation at low and 
high energies, respectively. XMM2 and SWIFT 2 have a similar but less
pronounced feature.  None of the other spectra appear to have two spectral
peaks. This is especially intriguing for SWIFT 1, as this has very similar
overall flux to XMM1 
yet is plainly different in shape, with more flux between 1 -- 2~keV filling
in the inflection, and less flux below 0.7 keV.  This difference could instead
be due to increased absorption in SWIFT 1 rather than a real spectral change
for the constant absorption column assumed here.
However, the low energy spectral shape of SWIFT 1 is similar to that of the
other SWIFT spectra, so it seems more likely that there is a real spectral
change between the SWIFT spectra and XMM1. This 
is not from the different instrument responses as XMM3 has a similar low
energy slope to SWIFT 2. Thus it seems most likely that there are real spectral
changes for little or no change in luminosity. 

At higher energies there is more variability. The high energy
rollover clearly indicates that the Comptonising corona has a lower
electron temperature in SWIFT 3 and 4 than in the lower luminosity
spectra (XMM1 -- 3 and SWIFT 1 and 2), but this trend is not continued to
the highest luminosity data (SWIFT 5), where the rollover again moves
up in energy to close to the edge of the bandpass.

We plot the model parameters in Fig. 4 as a function of the
unabsorbed, bolometric luminosity. 
The {\it XMM-Newton\/} data are plotted as filled stars: XMM1, XMM2, and XMM3,
from left to right, respectively.
The disc is consistent with having
the same temperature in all spectra ($\chi^2_\nu=3512/3461$).
Alternatively, it is consistent with having the same
normalisation in all spectra, but tying both temperature and
normalisation give a significantly worse fit at $\chi^2_\nu=3562/3475$.
This gives the almost constant low energy flux described
above.

The electron temperature is clearly changing, decreasing with
increasing luminosity through XMM1 -- 2 and SWIFT 1 -- 4, except for
XMM3 where we cannot have a good constraint on the upper limit of the
electron temperature, even worse than SWIFT 5 (and thus its bolometric flux
has likely been overestimated in Fig. 4 and Table 2). The similar
spectral slope of all the data in the 2 -- 4~keV range dominated by
Comptonisation requires a similar Compton $y$ parameter, defined as
$y=4 \tau^2 \theta=$~constant, where $\theta=kT_{\rm e}/m_{\rm e}
c^2$, so the lower temperature implies an increased optical depth. A
lower electron temperature also brings the rollover further into the
observed bandpass, so it is better constrained. Hence the optical
depth (derived from the observable parameters of spectral slope and
electron temperature) has correspondingly smaller errors also.
However, the highest luminosity spectrum does not follow this
trend. The electron temperature increases again, so the optical depth
drops.

\begin{figure}
 \includegraphics[angle=0,scale=0.45]{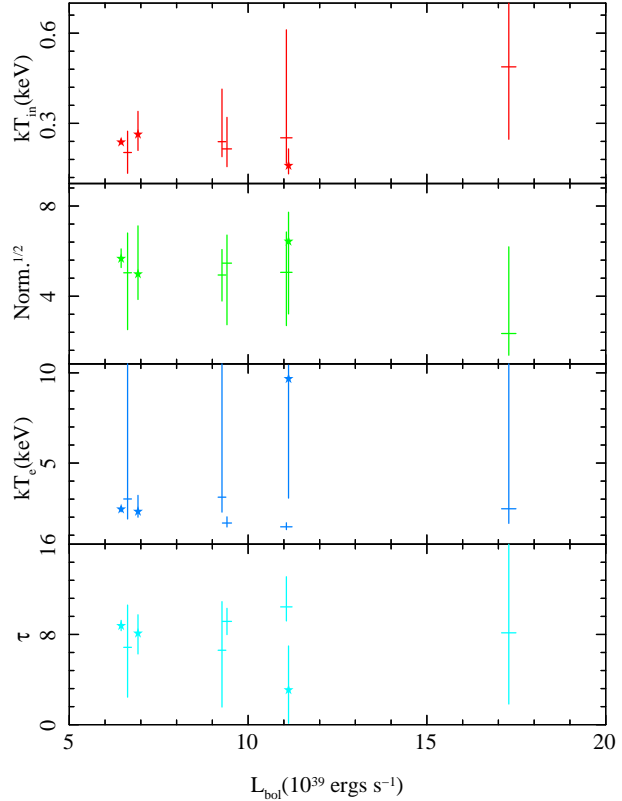}
 \caption{Dependence of the best fitting parameters on the bolometric
 luminosity.  From top to bottom: disc temperature $kT_{\rm in}$
 (keV); square root of {\tt diskbb} normalization (which is $\propto$ disc
 inner radius); electron temperature in the corona $kT_{\rm e}$ (keV);
 and the electron scattering optical depth $\tau$. Filled stars are
 {\it XMM-Newton\/} data: XMM1, XMM2, and XMM3 from left to right,
 respectively.}
\label{figure 4}
\end{figure}

\section{Discussion}

\citet{b7} suggested that the different spectral shapes seen in their
sample of ULXs could be set into a sequence such that the corona
became increasingly optically thick and lower temperature, extending
over more of the inner disc, as the mass accretion rate $\dot{M}$
increased beyond Eddington ($\dot{M}_{Edd}$). Plausibly these changes
could be caused by the increasingly strong winds which are expected to
be driven by such super-Eddington flows. At even higher
$\dot{M}/\dot{M}_{Edd}$ the source becomes embedded in an expanding
photosphere, lowering the temperature still further \citep{b44,b45}.

Our results show some support for this picture. Figs. 3 and 4 show
that the electron temperature and optical depth in the corona
generally decrease with increasing luminosity except for the brightest
spectrum.  However, they also highlight the fact that the spectral
variability does not uniquely track the observed luminosity. SWIFT 1
and XMM1 have very similar fluxes, but their spectra are subtly
different.  Such lack of one-to-one correspondence between spectral
shape and luminosity is a well-known feature of BHBs for the
transition between the low/hard and high/soft states. However, this
hysteresis in BHBs is most probably connected to non-equilibrium
effects caused by the rapid rise in mass accretion rate in the
transient outbursts of these systems \citep{b103}, whereas most ULX
are persistent sources.  BHBs can also however show another type of
non-unique relation of spectral shape to luminosity, where the
strongly Comptonised very high state can be seen at the same
luminosity as a standard disc dominated state \citep[and the
references therein]{b46}. It is not at all clear what determines which
of these high mass accretion rate states is shown at any time. The
difference between SWIFT 1 and XMM1 may then be connected to the same
(unknown) mechanism. Alternatively it may be connected to the
additional complexity expected from super-Eddington flows.  Numerical
simulations show that such flows are highly dynamical (Ohsuga 2006;
2007; 2009; Kawashima et al. 2009; Takeuchi, Mineshige \& Ohsuga
2009). Hence changes in the disc/wind geometry along the line of sight
could result in complex changes in the observed flux which do not
reflect the intrinsic luminosity of the source.  However, the complex
structure in accretion flow shown by the
numerical simulations mentioned above may have been smoothed out within
the observation timescale of our present study (of the order of days).
Alternatively, we may consider an intermittent outflow
(e.g. Kato, Mineshige, \& Shibata 2004). 

It is interesting to compare the variability behaviour of Holmberg IX
X-1 to other ULXs modelled with similar cool disc plus Comptonisation
models. Specifically, \cite{b104} explored this for the ULX NGC 5204
X-1 based on a {\it Chandra\/} monitoring programme and {\it
XMM-Newton\/} data, and \cite{b101} use this model to compare {\it
XMM-Newton\/} data for several ULXs over all available epochs with
sufficient data quality.  Interestingly, these works show that IC 342
X-1 behaves similarly to Holmberg IX X-1; its coronal electron
temperature appears to drop as it gains in luminosity.  In contrast,
both NGC 5204 X-1 and Holmberg II X-1 behave differently.  In these
ULXs the electron temperature increases with increasing luminosity,
with a corresponding drop in optical depth. Perhaps most
interestingly, there appears to be a correlation between these
spectral changes and the position of the sources in the spectral
sequence for ULXs shown as Fig. 8 of Gladstone et al. (2009).  Both
Holmberg IX X-1 and IC 342 X-1 appear in the central portion of the
sequence, where the spectra appear with a strong inflection and the
energy density peaks in the higher energy, optically thick,
Comptonised spectral component. Holmberg II X-1 and NGC 5204 X-1, on
the other hand, appear further along the sequence, where the energy
density peaks in the softer part of the spectrum and Gladstone et
al. (2009) hypothesize that the spectra are becoming
photosphere-dominated due to the increasingly important effects of a
strong wind.  This apparent correlation between spectral shape and
variability data supports the conclusion of Gladstone et al. (2009)
that different physical processes underly the two distinct spectral
shapes.  However, we emphasize that this tentative result is concluded
from the current, sparse data available for ULX spectral variability
studies; clearly new dedicated observational programmes are required
to investigate these phenomena in more detail.

\section{Conclusions}

We use the multiple {\it Swift\/} observations of Holmberg IX X-1,
co-added in different intensity ranges, together with a long {\it
XMM-Newton\/} observation of this source, to make the most sensitive
study to date of spectral variability in any individual ULX. We show
that the variability is subtle and rather complex, such that it is not
well characterised by a single hardness ratio. Instead, we do full
spectral fitting with a disc plus thermal Comptonisation model. 
There is a general trend in Holmberg IX X-1 for the Comptonising
corona to decrease in temperature and increase in optical depth as the
source luminosity increases. This supports the suggestion of
\citet{b7} that an increase in mass loss in winds from the inner disc as
the source becomes more super-Eddington leads to increased material in
the corona, thereby increasing the optical depth. The acceleration
mechanism is then required to share the coronal power across more
particles, so the temperature drops.  We also note that there appears
to be a tentative correlation between the observed changes in the
coronal parameters with luminosity, and the type of spectrum
displayed, over the small number of ULXs with sufficient quality data
to begin to explore spectral variability.

However, the data for Holmberg IX X-1 also show more complex
behaviour. It is clear that the spectral shape is not uniquely
determined by the observed luminosity. At the same luminosity the
spectra can look subtly different, and at different luminosities the
spectral shape can be the same. It may indicate that some fraction of
the observed variability is due to time dependent line of sight column
changes rather than intrinsic changes. Alternatively, this may be
related to the equally puzzling lack of one-to-one correspondence
between spectral shape and luminosity seen in the BHBs at high (but
apparently sub-Eddington) mass accretion rates where the source can
show either high/soft or very high state at the same luminosity.
Whatever the reason, this severely complicates any attempt to
understand the origin of the spectral variability.  It is therefore
very evident that if we are to make progress in understanding ULX
spectral variability, and the physical processes that underly these
behaviours, then the acquisition of new observational data composed of
good quality ULX spectra over a range of timescales and source fluxes
is absolutely imperative.

\section*{Acknowledgments}
We thank the referee for their useful comments, that helped improve this work.
KV thanks Chia-Ying Chiang for help with the {\it Swift\/} data reduction
and the Ministry of Education, Culture, Sports, Science and Technology
(MEXT) of Japan for the scholarship.
This work is supported in part by the Grant-in-Aid for the Global COE Program
"The Next Generation of Physics, Spun from Universality and Emergence"
from MEXT and by the Grant-in-Aid of MEXT (19340044, SM).
This work is in part based on observations obtained with {\it
XMM-Newton\/}, an ESA science mission with instruments and
contributions directly funded by ESA Member States and NASA.

\bsp

\label{lastpage}

\end{document}